\documentclass[submission, Proceedings]{SciPost}

\hypersetup{
    colorlinks,
    linkcolor={red!50!black},
    citecolor={blue!50!black},
    urlcolor={blue!80!black}
}

\usepackage[bitstream-charter]{mathdesign}
\DeclareSymbolFont{usualmathcal}{OMS}{cmsy}{m}{n}
\DeclareSymbolFontAlphabet{\mathcal}{usualmathcal}

\begin{document}

\begin{center}{\Large \textbf{
Studying hadronization at LHCb \\
}}\end{center}

\begin{center}
Sookhyun Lee\textsuperscript{1$\star$} on behalf of the LHCb collaboration,
\end{center}

\begin{center}
{\bf 1} University of Michigan, Ann Arbor, MI USA
\\
* sookhyun@umich.edu
\end{center}

\begin{center}
\today
\end{center}


\definecolor{palegray}{gray}{0.95}
\begin{center}
\colorbox{palegray}{
  \begin{tabular}{rr}
  \begin{minipage}{0.1\textwidth}
    \includegraphics[width=22mm]{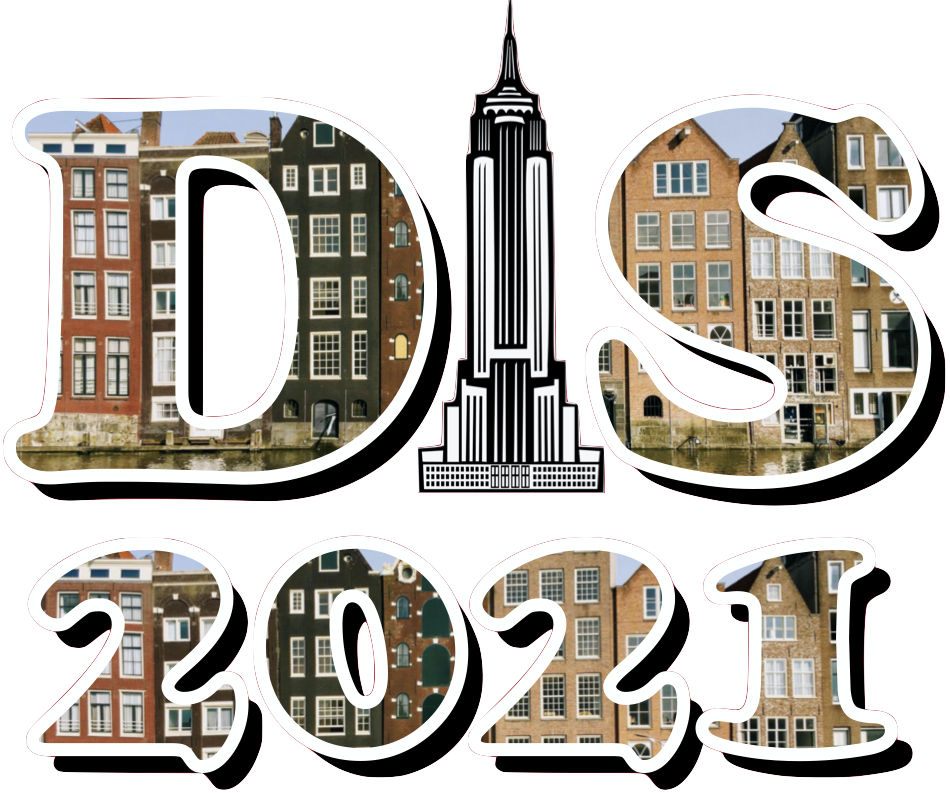}
  \end{minipage}
  &
  \begin{minipage}{0.75\textwidth}
    \begin{center}
    {\it Proceedings for the XXVIII International Workshop\\ on Deep-Inelastic Scattering and
Related Subjects,}\\
    {\it Stony Brook University, New York, USA, 12-16 April 2021} \\
    \doi{10.21468/SciPostPhysProc.?}\\
    \end{center}
  \end{minipage}
\end{tabular}
}
\end{center}

\section*{Abstract}
{\bf
LHCb is a fully instrumented forward spectrometer with particle identification and muon reconstruction covering the pseudorapidity ($\eta$) range from 2 to 5. Its full jet reconstruction capability makes the LHCb experiment a suitable venue to explore jet substructure observables, particularly formation of hadrons and heavy quarkonia resonances inside jets. This contribution presents a brief overview of ongoing research program and discusses recent results on non-identified charged hadron distributions in Z-tagged jets and charmonium distributions within jets. Future works towards furthering the knowledge of hadronizion are also discussed.

}


\section{Introduction}
\label{sec:intro}
Three dimensional imaging of the nucleon structure has gained enormous interests over the past decade in anticipation of forthcoming data at the Electron-Ion Collider. As introducing transverse momentum dependent (TMD) motion of partons inside the nucleon and its interactions with fundamental properties of the nucleon explained one of the greatest mysteries, e.g. large transverse single spin asymmetries (TSSA) measured in pp collisions, adding an additional dimension to the description of the formation of hadrons is only natural and will enrich the hadronic physics. On the other hand, with the success of Soft-Collinear Effective Theory (SCET) that allows solving multi-scale problems in a re-factorized manner, various observables probing substructure of hadro-produced jets have become calculable retaining predictive power of pQCD. Measurements presented here and other planned measurements involve looking inside jets for this reason.  

\section{Hadronization of charged particles in jets}
\label{sec:chhad}

In standard collinear perturbative QCD (pQCD) factorization framework, hadronization of charged particles is described by collinear fragmentation functions (FFs), denoted as $D_{c}^{h}(z, \mu)$, where $z$ is the momentum fraction of an outgoing parton $c$ carried by a hadron $h$ (see~\cite{Metz_2016} for a review). In this framework, single inclusive hadron production in $p$$p$ collisions factorizes into the the short distance hard scattering of partons and the long distance dynamics described by collinear FFs and collinear parton distribution functions (PDFs). The latter, denoted by $f_{a/p} (x, \mu)$, parameterize the momentum fraction $x$ of an incoming proton carried by a parton $a$ taking a part in hard scattering processes. The FFs and PDFs are non-perturbative (NP) objects that can only be measured in experiments, in contrast to the single inclusive jet production that is perturbatively calculable in pQCD. Jet fragmentation functions (JFFs) are experimental observables that measure the longitudinal momentum fraction of a jet carried by hadrons inside the jets. Within the SCET framework, the JFFs are defined in such a way that they can probe the collinear FFs. Analogously, TMD JFFs can access standard definition of TMD FFs measured in e+e- and SIDIS~\cite{Collins:2011zzd}. Theoretical development of this kind is crucial in the following sense. TMD factorization breaking is predicted for jet production in p+p~\cite{Rogers:2010dm}, in which case non-perturbative TMD distribution and fragmentation functions can no longer be identified as universal objects in different collision processes. As such, interpretation of these measurements has to be carried out with caution and test of factorization breaking and universality of non-perturbative FFs across different collision systems are indispensible. 
LHCb data provide constraints on these NP physics objects in kinematic regimes complementary to those accessible at central pseudorapidities, e.g. a region of $x$ smaller than $10^{-4}$ for TMD PDFs. The Drell Yan (DY) measurements with the $Z \rightarrow \mu\mu$ channel \cite{LHCb:2015okr,LHCb:2015mad,LHCb:2016fbk} at LHCb have been used in a recent global analysis \cite{Scimemi:2019cmh} of TMDs. The DY angular distributions data affected by Boer-Mulders TMD PDF are also expected to constrain TMD PDFs and test their universality. 
 

The unpolarized TMD FFs depend on $z$, $j_T$ and theory scale $\mu$, i.e. $\hat{D}_i^h (z, j_T, \mu)$, therefore our main observables naturally involve $z$ and $j_T$, and additionally $r$, experimentally defined in Eq.~\ref{eq:vars} and also found in Ref.~\cite{LHCb-PAPER-2019-012}.\\

\begin{equation}
z = \frac{p_{h} \cdot p_{jet}}{p^2_{jet}}, \qquad j_T = \frac{| p_{h} \times p_{jet} |} {p_{jet}} [GeV/c], \qquad r = \sqrt{d\phi^2_{h-jet} + d\eta^2_{h-jet}}
\label{eq:vars}
\end{equation}

where $z$ is the longitudinal momentum fraction of jet momentum carried by a hadron, $j_T$ is the transverse component of hadron momentum with respect to the jet axis and $r$ is the radial distance of a hadron from the jet axis.

The dominant hard partonic process for Z+jet production in the LHCb acceptance is $q g \rightarrow Z q$.  Z-tagged jets at LHCb are therefore predominantly initiated by quarks and enhance sensitivity to the quark TMD FFs. The asymmetric parton momentum fractions of incoming protons required for both Z and jet to be in the forward region further enriches the data with jets initiated by light valence quarks with a large x. 

The jet fragmentation function measurements of which results are being presented here are listed in Eq.~\ref{eq:obs}~\cite{Kang-tmdff-2019}, where the phase space $\mathcal{PS} = d\eta_{Z}d\eta_{jet}dp_Td^2\mathbf{q_{T}}$, where $\mathbf{q_T}$ is the vector sum of the transverse momenta of the Z boson and the jet, and $2 \mathbf{p_T}$ the difference between the two. 

\begin{equation}
F(z_h) = \frac{d\sigma}{d\mathcal{PS} \; dz_h}/\frac{d\sigma}{d\mathcal{PS}} ,\qquad \int dz_h \;f(z_h,  j_T) = \frac{d\sigma}{d\mathcal{PS} \; dj_T}/\frac{d\sigma}{d\mathcal{PS}} ,
\label{eq:obs}
\end{equation}
\begin{equation*}
G(r) = \frac{d\sigma}{d\mathcal{PS} \; dr} /\frac{d\sigma}{d\mathcal{PS}} , \qquad \textrm{where}\quad h : \pi^{\pm} \;\; \textrm{or} \;\; K^{\pm} \;\; \textrm{or} \;\; p^{\pm}.
\end{equation*}

Note that the second equation is equivalent to the TMD JFF, denoted by $f(z_h, j_T)$, integrated over $z_h$. Experimentally, these quantities can be expressed as shown in Eq.~\ref{eq:obsexp}. 

\begin{equation}
F = \frac{1}{N_{Z+jet}}\frac{dN_h}{dz_h}, \qquad \int dz_h f= \frac{1}{N_{Z+jet}}\frac{dN_h}{dj_T}, \qquad G = \frac{1}{N_{Z+jet}}\frac{dN_h}{dr}
\label{eq:obsexp}
\end{equation}



\subsection{Results}

\begin{figure}[h]
\centering
\includegraphics[width=0.325\textwidth]{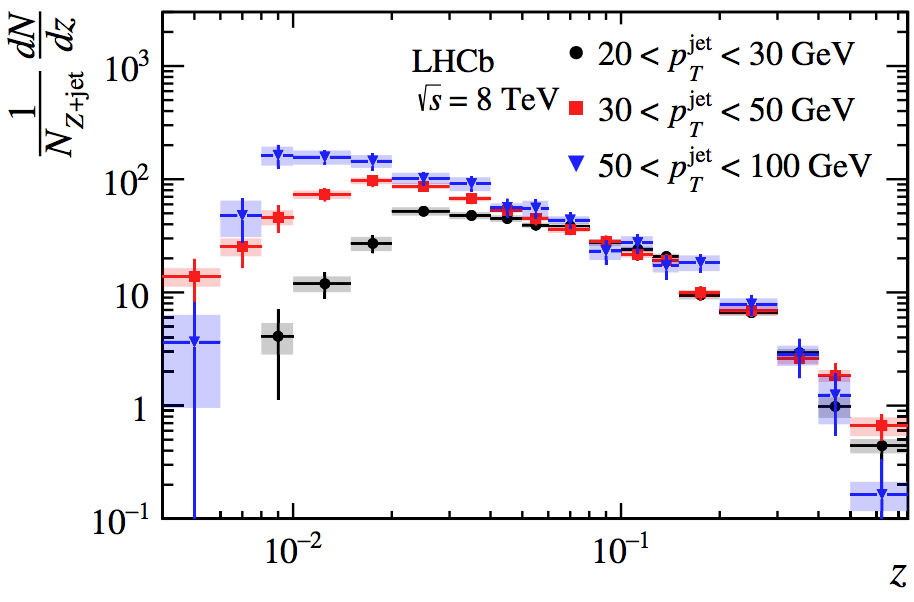}\includegraphics[width=0.32\textwidth]{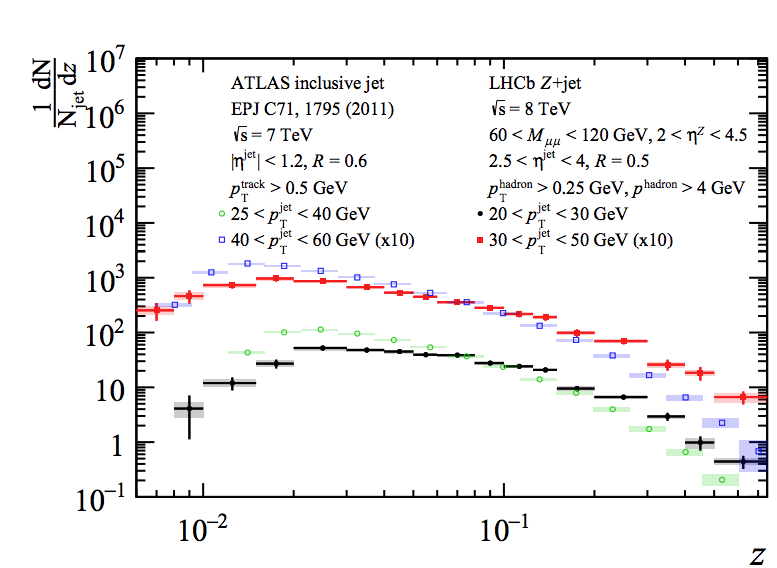}\includegraphics[width=0.315\textwidth]{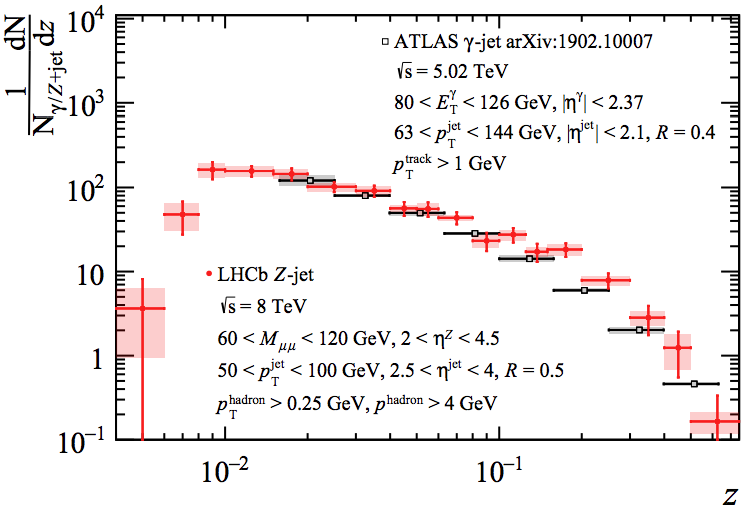}
\caption{Measured $z$ distributions at $\sqrt{s}$=8~TeV (left) and comparisons with ATLAS inclusive jets (middle) and ATLAS $\gamma$-jets (right)~\cite{LHCb-PAPER-2019-012}. }
\label{zdist}
\end{figure}

Measured longitudinal momentum distributions $z$ are shown in Fig.~\ref{zdist}. In the low $z$ region less than 0.04, a humped-back structure that is attributed to color-coherence effects compatible with parton-hadron duality~\cite{Azimov:1985by} is seen. More energy available to produce hadrons in the jets from harder scattering explains the increase of the height of the hump with jet $p_T$. At mid-to-high $z$, between 0.04 and 0.4, the fragmentation of a jet into hadrons is governed by the scaling behavior. The highest $z$ region is experimentally challenging due to limited statistics and degrading reconstruction efficiency. Theoretically, the presence of large logarithms that have to be re-summed with good precision adds to the challenge. Comparing the collinear JFF results of the LHCb $Z$+jet to the ATLAS single inclusive jet, an obvious difference seen in the shape of the distributions, an excess at higher $z$ and a depletion at lower $z$, is as expected based on the gluon dominance in the single inclusive jet production at LHC. A consistency between the LHCb $Z$+jet and the ATLAS $\gamma$+jet results confirms this interpretation.     

\begin{figure}[h]
\centering
\includegraphics[width=0.41\textwidth]{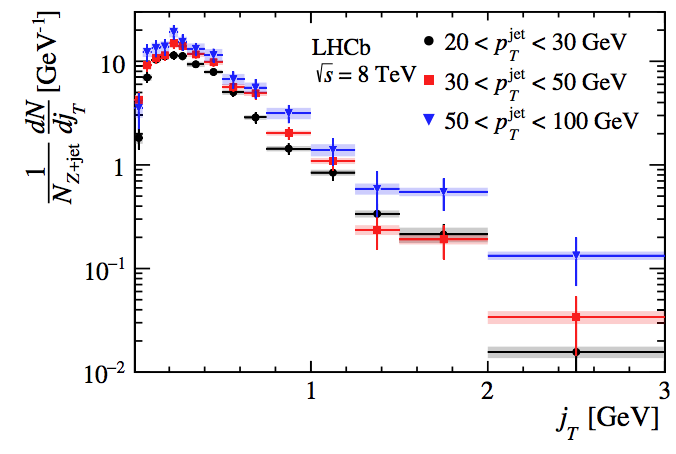}\includegraphics[width=0.42\textwidth]{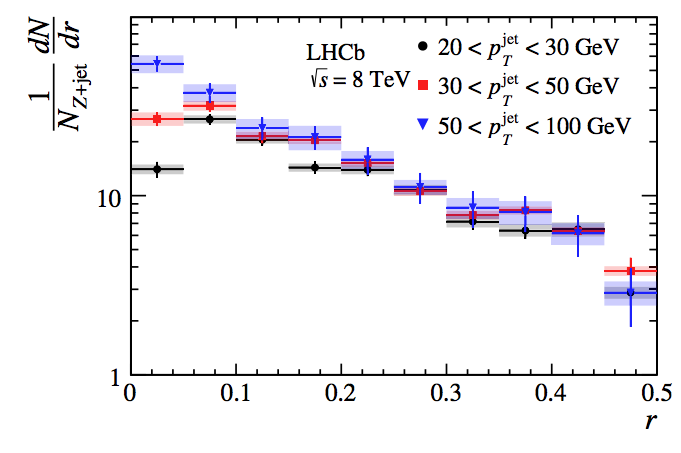}
\caption{Measured $j_T$ distributions (left) and $r$ distributions (right) at $\sqrt{s}$=8~TeV~\cite{LHCb-PAPER-2019-012}. }
\label{jt_r}
\end{figure}

Measured $j_T$ and radial distributions are shown in Fig.~\ref{jt_r}. The shape of the $j_T$ distributions does not show strong dependence on the $p_T$ of the jets, while harder jets tend to produce more particles in the collinear direction of the jets. 

The results are in good agreement with the TMD analysis performed in Ref.~\cite{Kang-tmdff-2019} in regions where perturbative calculations lack large logarithms that need to be resummed with high precision. This is promising for further studies to be published in the future.

\subsection{Future works}

Fully differential TMD JFF measurements, i.e. with respect to $j_T$ and $z_h$ are being studied at LHCb along with identified charged hadron distributions. Identified charged hadron distributions provide additional information on the valence versus sea quark's role initiating the parton shower that leads to the creation of jets. The corresponding observable is shown in Eq.~\ref{eq:obsnew}. When measured at a different beam energy, the observables can provide insights into the universality of the JFFs. 

\begin{equation}
f(z_h, j_T) = \frac{d\sigma}{d\mathcal{PS} dz_h dj_T}/\frac{d\sigma}{d\mathcal{PS}}, \quad \textrm{where}\quad h = \textrm{identified} \; \pi^{\pm}, K^{\pm}, p^{\pm}.
\label{eq:obsnew}
\end{equation}

Measurements that are sensitive to electromagnetic charge such as positive-to-negative charge ratio or jet charge~\cite{Kang:2020jc} can provide statistical sensitivities to different parton flavors initiating jets. Lastly, similar properties in heavy quark initiated jets are also being investigated to compare with light-quark initiated jets. 





\section{Formation of quarkonia in jets}
\label{sec:quarkonia}

Production mechanisms of heavy quarkonia are not well understood. LHCb published a slew of results on charmonium/bottomonium cross sections and polarizations~\cite{LHCb:2015log,LHCb:2015foc,LHCb:2018yzj,LHCb:2017scf,LHCb:2013izl}. Non-relativistic QCD (NRQCD), the most common approach to this problem, is generally successful in describing quarkonuia productions across different experiments. In this framework, the long distance matrix elements (LDMEs) that describe the nonperturbative transition of the $c\bar{c}$ pair in a state of definite color and angular momentum quantum number into a final state containing $J/\psi$ are extracted from data. The challenge is that almost all of the world data on quarkonia polarization contradict the high transverse polarization predicted from NRQCD using extracted LDMEs. An alternative approach to understanding quarkonium production mechanism is to measure quarkonium in jets. The usual observable discussed in the previous section is to measure is the collinear JFF of quarkonia, i.e. longitudinal momentum fraction of the jets carried by quarkonia $z = p_T^{J/\psi}/p_T^{jet} $. For the first time, LHCb measured this for charmonium, prompt as well as from $b$-hadron decay~\cite{LHCb:2017llq}. 

\subsection{Results}

\begin{figure}[h]
\centering
\includegraphics[width=0.9\textwidth]{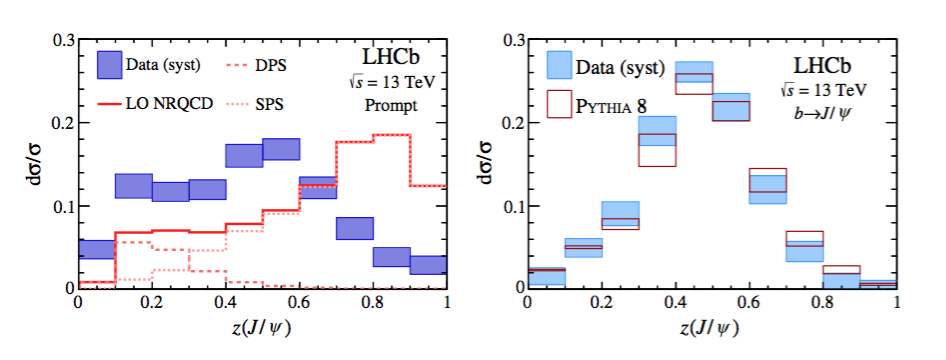}
\caption{Measured $z$ distributions for prompt $J/\psi$ (left) and from $b$-decay (right) at $\sqrt{s}$=13~TeV~\cite{LHCb:2017llq}. }
\label{charm}
\end{figure}

Measured  $z$ for prompt $J/\psi$ and from $b$-hadron decay are shown in Fig.~\ref{charm}. The data are compared to the leading-order (LO) NRQCD-based prediction as implemented in PYTHIA8. This prediction used LDMEs determined empirically both for color-octet and color-singlet intermediate states. While $b$ quark fragmentation into a jet with a $b$-hadron that subsequently decays into a charmonium state is fairly well described by the LO NRQCD predictions, the prompt $J/\psi$ counterpart in data takes much smaller momentum fraction of jets within which they are created than the PYTHIA8 predictions.

To reconcile the discrepancy between the data and the Pythia8 predictions, Ref.~\cite{Bain:2017wvk} performed a careful analysis using NRQCD. It was noted that the parton shower of the heavy quark-antiquark pair produced in a color-octet state was treated as a color singlet particle splitting into quark-antiquark pair 2 $P_{qq} (z)$, of which probability peaks at $z$=1. At LHC, quarkonia production is predominantly by gluon fragmentation. In \cite{Bain:2017wvk}, two methods were introduced to address this point, namely gluon-fragmentation-improved-PYTHIA and fragmenting jet function (FJF). Given the multi-scale nature of the quarkonium production, the latter approach is an appropriate solution, extended from collinear FJF defined for hadron JFF measurements. Both methods capture the softer $z$ seen in the LHCb prompt $J/\psi$ data.

\subsection{Future works}

The $z$ measurements can be repeated for bottomonium. The $z$-dependent polarization of charmonium as well as bottomonium in jets are being measured at LHCb. More differential measurements of this kind are expected to provide a better handle on understanding and constraining different production mechanisms of quarkonia. Separation between heavy-quark and gluon fragmentation is also an experimentally challenging question of interest.



\section{Conclusion}
LHCb has successfully performed measurements studying properties/mechanisms of hadronization using jet substructure technique, charmonium for the first time, followed by non-identified hadrons in jets. A number of follow-up measurements are being carried out towards gaining insights into the flavor dependence of the quark that initiates jets that produce charged hadrons and the full TMD picture of the hadronization process. Quarkonium production mechanisms are still not well understood. Nonetheless, new jet substructure type of measurements opened up ways to test the current state-of-the-art theory and raised interesting questions to be answered. Other measurements being explored include transverse polarization of $\Lambda$ baryons towards accessing standard TMD FF and $Z$-hadron correlations for testing TMD factorization breaking.


\section*{Acknowledgements}

\paragraph{Funding information}
The author is supported by the U.S. National Science Foundation.
\bibliographystyle{SciPost_bibstyle} 
\bibliography{references.bib}

\nolinenumbers

\end{document}